\documentclass[sigconf]{acmart}
\usepackage{bm}
\usepackage{bbding}
\AtBeginDocument{%
  }

\copyrightyear{2023}
\acmYear{2023}
\setcopyright{acmlicensed}\acmConference[MM '23]{Proceedings of the 31st ACM International Conference on Multimedia}{October 29-November 3, 2023}{Ottawa, ON, Canada}
\acmBooktitle{Proceedings of the 31st ACM International Conference on Multimedia (MM '23), October 29-November 3, 2023, Ottawa, ON, Canada}
\acmPrice{15.00}
\acmDOI{10.1145/3581783.3613825}
\acmISBN{979-8-4007-0108-5/23/10}

\begin{document}

\title{Face-Driven Zero-Shot Voice Conversion with Memory-based Face-Voice Alignment}


\author{Zheng-Yan Sheng}
\email{zysheng@mail.ustc.edu.cn}
\affiliation{%
  \institution{University of Science and Technology of China}
  \country{}}

\author{Yang Ai}
\email{yangai@ustc.edu.cn}
\affiliation{%
  \institution{University of Science and Technology of China}
  \country{}}

\author{Yan-Nian Chen}
\email{yannianchen@mail.ustc.edu.cn}
\affiliation{%
  \institution{University of Science and Technology of China}
  \country{}}

\author{Zhen-Hua Ling}
\email{zhling@ustc.edu.cn}
\authornote{Corresponding author.}
\affiliation{%
  \institution{University of Science and Technology of China}
  \country{}}

\begin{abstract}
This paper presents a novel task, zero-shot voice conversion based on face images (zero-shot FaceVC), which aims at converting the voice characteristics of an utterance from any source speaker to a newly coming target speaker, solely relying on a single face image of the target speaker. To address this task, we propose a face-voice memory-based zero-shot FaceVC method. This method leverages a memory-based face-voice alignment module, in which slots act as the bridge to align these two modalities, allowing for the capture of voice characteristics from face images.  A mixed supervision strategy is also introduced to mitigate the long-standing issue of the inconsistency between training and inference phases for voice conversion tasks. To obtain speaker-independent content-related representations, we transfer the knowledge from a pretrained zero-shot voice conversion model to our zero-shot FaceVC model. Considering the differences between FaceVC and traditional voice conversion tasks, systematic subjective and objective metrics are designed to thoroughly evaluate the homogeneity, diversity and consistency of voice characteristics controlled by face images. Through extensive experiments, we demonstrate the superiority of our proposed method on  the zero-shot FaceVC task. Samples are presented on our demo website\footnote{
Source code and audio samples are available at 
\url{https://levent9.github.io/ZeroshotFaceVC-demo/}}.


\end{abstract}

\ccsdesc[500]{Information systems~Multimedia content creation}


\keywords{voice conversion, zero-shot, face-voice alignment}


\maketitle

\section{Introduction}
Voice conversion (VC) aims to convert the voice characteristics of a source speaker to a target speaker while keeping the linguistic content unchanged \cite{mohammadi2017overview, sisman2020overview}. It has potential applications in various fields, such as communication aids for the speech-impaired \cite{veaux2013towards}, speaker de-identification \cite{srivastava2020evaluating}, and dubbing\cite{gan2022iqdubbing}.  Zero-shot VC \cite{qian2019autovc, yuanimproving} is a specific VC task, which allows for the conversion of voices from any source speakers to a newly coming (i.e., unseen in the training data) target speaker using  only one reference utterance from the target speaker.
Zero-shot VC has gained much research attention in recent years considering its flexibility and less dependency on the amount of training data from target speakers. 

In addition to voice, face image is another modality that also carries information about individual identities. Some speaker characteristics, such as age and gender,  can be inferred from both voices and face images.  Previous studies \cite{mavica2013matching, kamachi2003putting,smith2016concordant} have presented evidences to support the notion that people can accurately identify an unfamiliar voice and a static image of the corresponding face belonging to the same person with greater than chance accuracy.  

Therefore, this paper presents a novel VC task named zero-shot voice conversion based on face images (i.e., zero-shot FaceVC). Instead of using  reference utterances, the task leverages a single face image from the unseen target speaker to convert the utterances from any source speakers. The objective of this task is to explore to what extent facial properties can serve as indicators of voice characteristics. If it is feasible to infer suitable voice characteristics from the face image of an unseen speaker, zero-shot FaceVC can hold immense potential in various applications. For instance, an editable virtual face can be used to produce a personalized voice  in virtual anchors, and 
automatic movie dubbing can generate voices more consistent  with characters' appearance. 

To the best of our knowledge, there are only  two existing studies \cite{kameoka2019crossmodal, lu2021face} on voice conversion based on face images (i.e., FaceVC). Different from the zero-shot scenario considered in  this paper, both of them explored the task of many-to-many FaceVC, where both source and target speakers were seen in the training set. In one study \cite{kameoka2019crossmodal}, the speaker embeddings utilized in the original VC model were replaced with face embeddings, which were estimated by reconstructing face images. 
Another study \cite{lu2021face} employed a three-stage training strategy, including face-voice reparameterization and facial-to-audio transformation, to achieve better performance. 

The most critical challenge in FaceVC is face-voice alignment, i.e., to derive corresponding voice representations given face representations.
In previous studies, face representations were estimated by either relying on the supervision of mel-spectrum reconstruction \cite{kameoka2019crossmodal} or minimizing the  mean square error (MSE) between speaker embeddings and face embeddings \cite{lu2021face}.
The former strategy \cite{kameoka2019crossmodal} can't be extented to zero-shot FaceVC since the recordings of target speakers are unavailable. The latter \cite{lu2021face} adopted the simple MSE loss, assuming that the distribution of voice representations given face representations was unimodal. 
Both of them failed to describe the complex mapping relationship between voice and face spaces, and can't fulfill the requirement of the zero-shot FaceVC task.

Therefore, this paper proposes a face-voice memory-based zero-shot FaceVC (FVMVC) method. In this method, a memory-based face-voice alignment (MFVA) module is developed  that utilizes trainable slots to quantize the common characteristics between face and voice spaces.
At the training stage, the slot values in MFVA are optimized by not only minimizing the reconstruction loss of speaker embeddings, but also reducing the  Kullback-Leibler divergence between the slot weight distributions in both spaces. At the inference stage, given a face image of an unseen target speaker, a recalled face embedding is calculated 
using the slot weights estimated from the reference image and the slot values in the voice space.

In addition, Zero-shot VC usually adopts  an auto-encoder framework \cite{qian2019autovc,lian2022robust, yuanimproving, wang21n_interspeech}, which suffers from the inconsistency between the training and inference phases. More specifically, the speaker representations and content representations are from the same speaker at the training stage, while they are from different speaker at the conversion stage. To mitigate this problem for zero-shot FaceVC, we propose a mixed supervision strategy, introducing a simple yet effective inter-speaker supervision in addition to the intra-speaker supervision in traditional auto-encoder frameworks. The inter-speaker supervision is achieved by creating pseudo-parallel training data using the speaker embeddings extracted from the recordings of another speaker in the training set. Besides, in order to obtain speaker-independent content representations, we initially pretrain a zero-shot VC model and transfer the knowledge from zero-shot VC to zero-shot FaceVC.

We have noticed that it should be impossible to recover the exact voice  of target speakers from only face images. Instead, we focus on three properties of the speech generated by zero-shot FaceVC. The first one is the homogeneity among the voice characteristics of the speech converted using different face images of the same target speaker. The second is the diversity of the voice characteristics converted using the face images of different target speakers. And the third is the consistency between the voice characteristics of the converted utterances and their corresponding face images in some important aspects, e.g., gender. 
Therefore, a series of subjective and objective metrics are designed in this paper to evaluate these properties mentioned above. 

In summary, our main contributions are as follows. 
First, we propose a new task named zero-shot voice conversion based on face images (zero-shot FaceVC). Second, we propose a face-voice memory-based zero-shot FaceVC (FVMVC) method for this task, which contains a memory-based face-voice alignment module, a mixed supervision strategy and zero-shot VC pretraining.  Third, we design a series of metrics to evaluate the proposed task and conduct extensive experiments to demonstrate the effectiveness of our proposed method.


\section{Related Work}
\subsection{Voice Conversion}
VC is a task that automatically converts the speech from a source speaker to a speech sound like being spoken by a target speaker while preserving the linguistic content \cite{mohammadi2017overview, sisman2020overview}.  This task can be categorized into two categories: parallel and non-parallel conditions.
Since parallel data are not always available, many non-parallel VC techniques have been proposed, including the methods based on variational auto-encoder (VAE) \cite{kameoka2019acvae,saito2018non}, generative adversarial network (GAN) \cite{kaneko2018cyclegan, kameoka2018stargan, wang2020one, lee2021voicemixer},  recognition-synthesis \cite{chen2022improving, mohammadi2019one, saito2018non} and disentanglement \cite{zhang2019non}. 

As a special case of non-parallel VC,  zero-shot VC has attracted widespread attention in recent years. 
Zero-shot VC methods usually follow auto-encoder frameworks, where the encoder extracts content and speaker representations from speech respectively, and the decoder reconstructs speech by combining the above representations.  Hence, speech representation disentanglement is  crucial for this task \cite{yang2022speech,wang21n_interspeech}. Recently, several zero-shot VC methods \cite{yuanimproving, wang21n_interspeech, yang2022speech} based on information theory have emerged, with the aim of disentangling the content-related and speaker identity-related information. IDE-VC \cite{yuanimproving} employed mutual information (MI) with speaker labels as supervision for disentanglement. VQMIVC \cite{wang21n_interspeech} combined vector quantization with contrastive predictive coding (VQCPC) \cite{van2020vector, baevski2019vq} and MI for fully unsupervised training.

\begin{figure*}[htb] 
  \includegraphics[width=\textwidth]{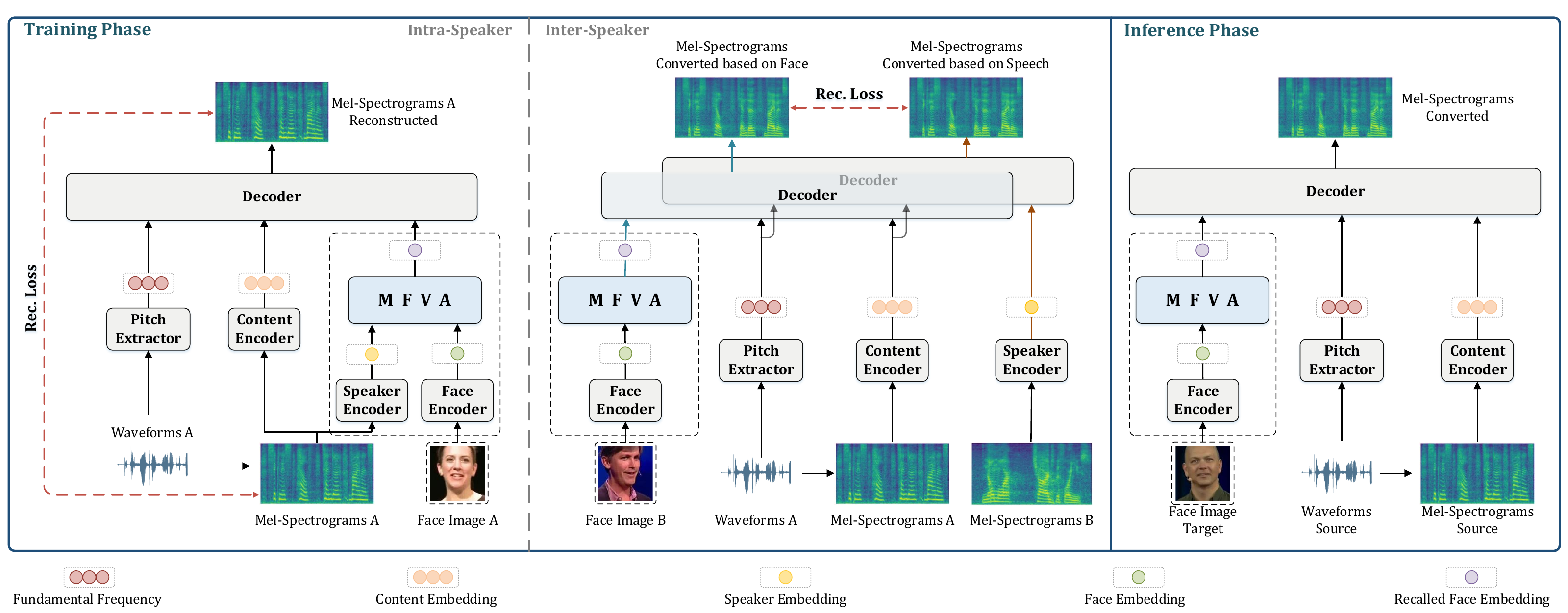}
  \caption{The overall flowchart of our proposed FVMVC, where Rec. Loss represents reconstruct loss. During the training phase, two pairs of utterances and the corresponding face images from speaker A and speaker B are used for training simultaneously. Speaker A is used for intra-speaker training and is selected as the source speaker for inter-speaker training, while speaker B is chosen as the target speaker for inter-speaker training. }
\label{fig1}
\end{figure*}

\subsection{Learning Voice-Face Association}
In recent years, learning the voice-face association has aroused the interest of researchers. As face and voice are inherently correlated, various cross-modal generation tasks involving both face and voice have been proposed, in addition to voice conversion supported by face images. Examples of such tasks include generating the talking face video from the audio \cite{park2022synctalkface, zhou2019talking, chen2019hierarchical, song2018talking, zhou2021pose}, synthesizing speech from the talking face images \cite{prajwal2020learning, kim2021lip, wang2022vcvts}, reconstructing the face image from the corresponding voice \cite{oh2019speech2face, wu2022cross, choiinference, wen2019face, bai2022speech} and synthesizing the speaker's voice with a face image during text-to-speech (FaceTTS) \cite{goto2020face2speech, yang2023does, pluster2021hearing, wang2022residual,wu2022speech,yang2022speaker}.

The most relevant task to voice conversion based on face images is  FaceTTS, as they both utilize face images to extract speaker identities for controlling voice characteristics. As far as we know, Face2Speech \cite{goto2020face2speech} was the first work to address FaceTTS, which pretrained a face encoder with the supervisedly generalized end-to-end (GE2E) \cite{wan2018generalized} loss and then replaced the speaker encoder with the face encoder in a multi-speaker TTS model. Following the Face2Speech framework, more elaborate
model structures and training strategies \cite{wang2022residual, wu2022speech, pluster2021hearing} have been proposed to promote the quality of synthetic speech. Recently, 3D face shapes and refined face  attributes have also been utilized to generate speech \cite{yang2023does,yang2022speaker}, which provided a referable approach to voice editing. However, the voice-face alignment in FaceTTS has yet to be explored. This paper elaborately designs a memory-based module for the alignment between these two modalities for zero-shot FaceVC, which can also be inserted into the FaceTTS framework for voice control.


\section{METHOD}
As shown in Figure \ref{fig1}, our proposed FVMVC follows the standard auto-encoder paradigm, consisting of a content encoder, a speaker encoder, a face encoder, a pitch extractor, a decoder, and a memory-based face-voice alignment (MFVA) module.  

During the inference phase, our proposed FVMVC utilizes three inputs, including a face image from the target speaker, together with the waveforms and the mel-spectrograms of the utterance to be converted from the source speaker. By processing the face image through a face encoder and an MFVA module sequentially, the voice characteristics representation based on the face image (i.e., the recalled face embedding) is obtained.  Similar to zero-shot VC, the mel-spectrograms of the source speaker provide a speaker-independent content representation, while waveforms are used for extracting normalized fundamental frequencies. All the representations mentioned above are ultimately sent into the decoder, which generates mel-spectrograms of the converted utterance. These mel-spectrograms are then converted to waveforms through the vocoder.  During the training phase, we incorporate speaker embeddings, which are extracted from mel-spectrograms via the speaker encoder,  to supervise the training of the MFVA module. 


\subsection{Memory-based Face-Voice Alignment}
\label{sec:memory}
The alignment between face and voice is intended to retrieve the corresponding speaker embedding when only a face image is available.  The retrieved speaker embedding from a face image is referred to as the recalled face embedding. 
We introduce the MFVA module to improve the modeling of voice-face alignment, thereby promoting the performance of zero-shot FaceVC.

\begin{figure}[h] 
  \includegraphics[width=0.45\textwidth]{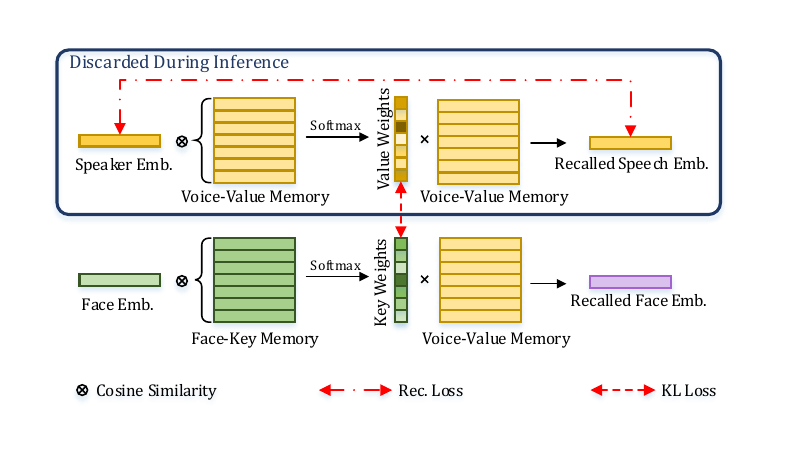}
  \caption{The architecture of the MFVA module, where Rec. Loss and KL Loss represent reconstruct loss and Kullback-Leibler divergence loss, respectively.}
\label{fig2}
\end{figure}

As shown in Figure \ref{fig2}, during the training phase, the MFVA module takes a pair of face embedding ${\bm{h}\in \mathbb{R} ^D}$ and speaker embedding ${\bm{s}\in \mathbb{R} ^D}$ as input, and generates a recalled face embedding ${\bm{\hat{h}}\in \mathbb{R} ^D}$for voice control, where the face embedding ${\bm{h}}$ and the speaker embedding ${\bm{s}}$ are extracted by the face encoder and the speaker encoder respectively, and $D$ represents the dimension of the projected face or speech embedding. MFVA is composed of a voice-value memory ${\bm{M}_{voice} = [\bm{m}_{v}^1, \bm{m}_{v}^2, \cdots,  \bm{m}_{v}^N ]^\intercal}$ ${\in \mathbb{R} ^{N \times D}}$ and a face-key memory ${\bm{M}_{face}=[\bm{m}_{f}^1, \bm{m}_{f}^2, \cdots, \bm{m}_{f}^N]^\intercal }$ ${\in \mathbb{R} ^{N \times D}}$, where $N$ denotes the number of the slots and $D$ is the dimension for each slot, which equals to the dimension of the projected speaker embedding or face embedding.  The training of the MFVA module contains two objectives, i.e.,  (1) storing sufficient voice characteristics information in voice-value memory, and (2) minimizing the distance between the distributions of two modalities.

\textbf{The sufficiency of the voice characteristics information}. Voice-value memory ${\bm{M}_{voice}}$ is made up of a bank of trainable slots ${\{\bm{m}_{v}^i}\}_{i=1}^N$, where $\bm{m}_{v}^i\in \mathbb{R} ^D$ is the $i$-th slot. The voice-value memory is designed to exclusively capture the voice-related information and expected to generate any voice.  Specifically, when we take a speaker embedding ${\bm{s}}$ as a query,  the attention weight between the query and each slot is computed with cosine similarity and softmax normalization function as follows,
\begin{equation}
w_{v}^i=softmax(\frac{ \bm{s}^\intercal \bm{m}_{v}^i}{\left\|\bm{s}\right\|_2\left\|\bm{m}_{v}^i\right\|_2 }),
\end{equation}
where $w_{v}^i$ represents the degree of relevance between the $i$-th slot $\bm{m}_{v}^i$ and the speaker embedding ${\bm{s}}$. Then we can get the attention weight vector ${\bm{w}_{voice}}=[w_{v}^1, w_{v}^2, \cdots, w_{v}^N]\in \mathbb{R} ^N$ by computing attention weight with all slots. In the end, we reconstruct the speaker embedding from all slots,
\begin{equation}
\hat{\bm{s}}= \bm{M}_{voice}^\intercal \bm{w}_{voice},
\end{equation}
and minimize the MSE between the input speaker embedding ${\bm{s}}$ and the recalled speaker embedding $\hat{\bm{s}}$, 
\begin{equation}
\mathcal{L}_{store}=\| \bm{s} - \hat{\bm{s}} \Vert _2 ^2.
\end{equation}
In this way,  the slots in the voice-value memory can be used as basis vectors for building the voice characteristics space, and various combinations of the slots can represent arbitrary voices.

\textbf{The alignment between voice and face space}. We utilize the slots as a streamlined bridge to map the face embedding onto the voice space. In detail, given the face embedding ${\bm{h}}$, we generate the recalled face embedding ${\hat{\bm{h}}}$ in a similar way as the recalled speaker embedding, where the attention weight is calculated with the face-key memory ${\bm{M}_{face}}$  and aligned with slots in voice-value memory ${\bm{M}_{voice}}$ as follows,
\begin{equation}
  w_{f}^i=softmax(\frac{\bm{h}^\intercal \bm{m}_{f}^i}{\left\|\bm{h}\right\|_2 \left\|\bm{m}_{f}^i\right\|_2 }),
\end{equation}
\begin{equation}
{\bm{w}_{face}}=[w_{f}^1, w_{f}^2, \cdots, w_{f}^N],
\end{equation}
\begin{equation}
  \hat{\bm{h}}=\bm{M}_{voice}^\intercal \bm{w}_{face},
\end{equation}
where $\bm{m}_{f}^i\in \mathbb{R} ^D$ the $i$-th trainable slot in the face-key memory ${\bm{M}_{face}}$.  Consequently, we combine the slots that are solely related to the voice to generate the recalled face embedding $\hat{\bm{h}}$. In addition, face images often contain various background noises and irrelevant information to the voice, such as shooting angle and image background. Voice-value memory ${M_{voice}}$ can impose an information bottleneck to remove these non-essential details.
In the end, we align the slot-weight distributions between two modalities using Kullback-Leibler divergence,
\begin{equation}
  \mathcal{L}_{align}=D_{KL}(\bm{w}_{voice}||\bm{w}_{face}).
\end{equation}
In this way,  during the inference phase, only given the face embedding as input, we can generate reasonable recalled face embedding to support the voice characteristics information for zero-shot FaceVC.

\subsection{Structures of Other Modules}
\label{sec:structure}
\textbf{Content Encoder} uses vector quantization (VQ) \cite{van2020vector} with contrastive predictive coding (CPC) \cite{baevski2019vq} to extract content embedding from mel-spectrograms, where VQ can be seen as an information bottleneck to remove inconsequential content information and CPC is used to explore the local structure of speech.

\noindent\textbf{Face Encoder} is used to extract the face embedding from the face image. We firstly leverage MTCNN \cite{zhang2016joint} for face detection on the original face image. Then we enhance the pretrained FaceNet \cite{schroff2015facenet} with a self-attention module to extract the facial feature as the face embedding.  During the zero-shot FaceVC training phase, only the self-attention module is jointly trained with other modules. 

\noindent\textbf{Speaker Encoder} takes in mel-spectrograms to generate the length-fixed speaker embedding.   The speaker encoder consists of two parts: \textit{Resemblyzer}\footnote{https://github.com/resemble-ai/Resemblyzer} and a self-attention module.  The \textit{Resemblyzer} is pretrained in the speaker verification task with GE2E loss function \cite{wan2018generalized}.
To better integrate the pretrained speaker verification network into the zero-shot VC network, we enhance the pretrained speaker verification network with a self-attention module, which is jointly trained with other modules during the zero-shot VC training phase.

\noindent\textbf{Pitch Extractor} extracts fundamental frequencies from input waveforms based on the period detection of the vocal fold vibration \cite{morise2009fast} and performs z-normalization for each utterance.  

\noindent\textbf{Decoder} maps the content representation, speaker-identity representation, and pitch representation into mel-spectrograms. The decoder mainly consists of convolutional blocks and two long short-term memory (LSTM) layers.

\subsection{Mixed Supervision Strategy}
\label{sec:mixed}
The mixed supervision strategy contains intra-speaker supervision and inter-speaker supervision. The left part in Figure \ref{fig1} shows the intra-speaker supervision following the traditional auto-encoder paradigm. During the training phase, we firstly encode the mel-spectrograms $\bm{X}_A$ of input utterance from Speaker $A$  into a speaker-independent content embedding ${\bm{c}_A}$, normalized fundamental frequencies ${\bm{f}_A}$  and a length-fixed speaker embedding ${\bm{s}_A}$. Meanwhile, we utilize the corresponding face image $\bm{Z}_A$  to get the face embedding ${\bm{h}_A}$. Then, we feed the face embedding $\bm{h}_A$ and the speaker embedding ${\bm{s}_A}$ into the MFVA module to get recalled face embedding $\hat{\bm{h}}_A$. In the end, the decoder $D$ maps the above representations to reconstruct the mel-spectrograms ${\hat{\bm{X}}_A=D(\bm{c}_A,\bm{\hat{h}}_A,\bm{f}_A)}$.  The decoder is jointly trained with the MFVA and the self-attention module in the face encoder by minimizing the following reconstruction loss,
\begin{equation}
\mathcal{L}_{Intra}=\|\hat{\bm{X}}_A - \bm{X}_A \Vert _2 ^2 + \|\hat{\bm{X}}_A - \bm{X}_A \Vert _1 .
\end{equation}

However, the training-inference inconsistency phenomenon exists when only adopting intra-speaker supervision, because the content embedding and the recalled face embedding are from the same speaker during the training phase while different during the inference phase.  We introduce a simple but efficient inter-speaker supervision strategy when the parallel corpus is unavailable. In this strategy, the voice converted by the target speaker embedding is used as the pseudo-parallel corpus of the face embedding for supervised training.  Specifically, as shown in the middle part of Figure \ref{fig1}, in addition to the speaker $A$, a pair of face image and mel-spectrograms of an extra speaker ${B}$  is as input, and the speaker embedding ${\bm{s}_B}$ and the recalled face embedding $\hat{\bm{h}}_B$ are obtained in the same way as inference phase.  Then speaker A is treated as the source speaker and speaker $B$  is viewed as the target speaker to convert voice as follows,
\begin{equation}
\bm{X}_{speech} = D(\bm{c}_A, \bm{s}_B, \bm{f}_A),
\end{equation}
\begin{equation}
\bm{X}_{face}=D(\bm{c}_A, \hat{\bm{h}}_B, \bm{f}_A),
\end{equation}
where $\bm{X}_{speech}$ refers to the converted mel-spectrograms that are supported by the speaker embedding $\bm{s}_B$,  and $\bm{X}_{face}$ denotes the converted mel-spectrograms obtained using the recalled face embedding $\hat{\bm{h}}_B$. Then, we optimize the MFVA by minimizing the reconstruction loss,
\begin{equation}
\mathcal{L}_{Inter}=\|\bm{X}_{speech} - \bm{X}_{face} \Vert _2 ^2 + \|\bm{X}_{speech} - \bm{X}_{face} \Vert _1 .
\end{equation}
In summary, the final loss function during the training phase of zero-shot FaceVC is as follows,
\begin{equation} \label{eq12}
  \mathcal{L}=\lambda _ 1   \mathcal{L}_{store} + \lambda _ 2 \mathcal{L}_{align} + \lambda _ 3  \mathcal{L}_{Inter} + \mathcal{L}_{Intra}, 
\end{equation}
where $\lambda _ 1, \lambda _ 2$, and $\lambda _ 3$ are constant weights to control how the importance of each term, and $ \mathcal{L}_{store}$ and $\mathcal{L}_{align}$  are described in Section 3.1.

\subsection{Pretraining Strategy}
\label{sec:pretrain}
It is widely recognized that the content encoder may encode the speaker identity-related information.  Only when the speaker and content representations are disentangled,  the voice characteristic of the utterance can be  converted by changing the speaker-identity representation \cite{qian2019autovc, yuanimproving}.  Hence speech representation disentanglement is a critical factor that significantly impacts the performance of the zero-shot FaceVC.  Taking this into considerations, we first pretrain a zero-shot VC model and then transfer the content encoder, speaker encoder,  and decoder to the zero-shot FaceVC for better performance. Specifically, during the training phase of the zero-shot FaceVC, the pretrained content encoder and the pretrained speaker encoder are fixed, while the pretrained decoder is further optimized with other modules.  

In order to achieve speech representation disentanglement, mutual information (MI) is introduced to evaluate the dependency between different representations.
We minimize the MI between content embeddings, speaker embeddings and fundamental frequencies utilizing the variational contrastive log-ratio upper bound (vCLUB) \cite{cheng2020club} during the training phase of zero-shot VC. Except for the MI loss, reconstruction loss, InfoNCE loss \cite{oord2018representation} and VQ loss \cite{van2020vector} are used to optimize the zero-shot VC model.  For more information on these functions, please refer to the VQMIVC \cite{wang21n_interspeech} paper.

\section{Experiments}
\subsection{Datasets}
Zero-shot FaceVC tasks place high demands on datasets, requiring not only a large volume of background-noise-free speech from various speakers but also clear face images of the corresponding individuals. Based on the above considerations, 
we conducted the experiments on the LRS3-TED \cite{afouras2018lrs3} dataset to evaluate the zero-shot FaceVC task.  This dataset includes 5,594 TED and TEDx talks in English, 
totaling over 400 hours of video content. The cropped face tracks in the video are provided at a resolution of 224×224 with a frame rate of 25 frames per second. The audio tracks are also available in a single-channel 16-bit 16 kHz format. In this dataset, the duration of speech from different speakers follows a long-tail distribution, which means that the majority of speakers have only a small number of utterances. To address the issue of uneven video distribution among speakers, we opted to use the top 200 speakers with the highest number of videos as our training set and validation set. During inference, we randomly selected a total of 12 newly coming speakers not in the training and validation sets: 8 target speakers (4 female and 4 male) and 4 source speakers (2 female and 2 male) for evaluation. 


\begin{table*}[]
\caption{Objective and subjective evaluation results of comparison systems. The definitions of all metrics can be found in Section \ref{metrics}.}
\label{tab:compare}
\begin{tabular}{lccccccc}
\toprule
\multicolumn{1}{c}{\textbf{}}       & \multicolumn{2}{c}{Homogeneity}   & \multicolumn{2}{c}{Diversity}     & \multicolumn{2}{c}{Consistency}   & Quality         \\ \cline{2-8} 

\multicolumn{1}{c}{\textbf{Method}} & SHR $\uparrow$            & SHO $\uparrow$            & SDR $\downarrow$            & SDO $\downarrow$            & GA $\uparrow$             & MOS-FVC $\uparrow$        & MOS-SN $\uparrow$   \\ \midrule
\textbf{Ground Truth}               & 0.8245          & 1.0000          & 0.5524          & 0.5524          & 1.0000          & 3.7042          & 4.2183          \\
\textbf{SpeechVC}                   & 0.7267          & 0.8229          & 0.5890          & 0.6408          & 0.9895          & 3.5917          & 3.6022          \\ \midrule
\textbf{Auto-FaceVC} \cite{lu2021face}                 & 0.7186          & 0.8132          & 0.6351          & 0.7042          & 0.9239          & 3.4289          & 3.5969          \\
\textbf{attentionCVAE} \cite{yang2022speaker}               & 0.7153          & 0.8081          & 0.6874          & 0.7789          & 0.9166          & 3.4292          & 3.5982          \\
\textbf{FVMVC}                      & \textbf{0.7313} & \textbf{0.8692} & \textbf{0.6188} & \textbf{0.6781} & \textbf{0.9791} & \textbf{3.5417} & \textbf{3.5993} \\ \bottomrule
\end{tabular}
\end{table*}

\subsection{Implementation Details}
To extract acoustic features, we first extracted the audio from the video clips with the FFmpeg tool \cite{tomar2006converting}.  Then 80-dim mel-spectrograms and normalized fundamental frequencies were calculated with a 25ms Hanning window, a 10ms frame shift and a 400-point short-time Fourier transform (STFT). We extracted a 512-dimensional face embedding from each frame of the video using MTCNN \cite{zhang2016joint} and FaceNet \cite{schroff2015facenet} subsequently. The dimension of face embeddings was projected into ${D=256}$, which was the same as the dimension of speaker embeddings and the slot dimension. The number of slots in MFVA was $N=96$. We applied the pretrained Parallel WaveGAN (PWG) vocoder \cite{yamamoto2020parallel} to convert the mel-spectrograms to waveforms.

For the pretraining strategy, the zero-shot VC model was trained for 1000 epochs using a batch size of 256. The mini-batch Adam optimizer was initialized with a learning rate of 1e-6 and was warmed up to 1e-3 after 2000 iterations. The learning rate was then decayed by a factor of 0.5 at epochs 300, 400, and 500. At the zero-shot FaceVC training stage, the model was updated for 2000 epochs using a batch size of 256. Similar to pretraining the zero-shot VC model, the learning rate of the Adam optimizer was initialized as 1e-6 and warmed up to 2.5e-4 after 3000 iterations. The learning rate was then decayed by a factor of 0.5 at epochs 800, 1200, and 1600. The constant weights ${\lambda _ 1}$, ${\lambda _ 2}$, and ${\lambda _ 3}$ in Equation \ref{eq12} were respectively set to be 1, 10, and 0.2.

\subsection{Comparison systems}
As we are the first to attempt the zero-shot FaceVC task and there are no existing comparable methods, we compared our proposed method with the following systems to evaluate its performance:

(1) \textbf{Ground Truth}: This method transferred the natural mel-spectrograms of target speakers to waveforms using the pretrained PWG vocoder. Since there are no parallel utterances between source and target speakers, the ground truth results can't be compared directly with the converted results, and are just used for indicating the upperbound of various metrics. 

(2) \textbf{SpeechVC}: This is our pretrained zero-shot VC model using natural reference utterances of target speakers for inference. 

(3) \textbf{Auto-FaceVC} \cite{lu2021face}: This method originally adopted AutoVC \cite{qian2019autovc} as the backbone. To better adapt the model to the zero-shot FaceVC task, we altered its backbone AutoVC to VQMIVC while preserving its original training strategy.

(4) \textbf{attentionCVAE} \cite{yang2022speaker}: This method used the face attributes to control the voice characteristics in the multi-speaker text-to-speech task. To adapt it to our zero-shot FaceVC task, we inserted its face attributes-based voice control module to replace the face encoder and the MFVA module in our proposed model.

 \subsection{Metrics}
 \label{metrics}
We developed several objective metrics to evaluate the homogeneity, diversity and consistency of the converted voice. The underlying motivation for measuring homogeneity is that the zero-shot FaceVC system should generate homogeneous voice characteristics with different face images from the same target speaker, regardless of the image's shooting angle and background. We applied the well-known open-source speaker verification toolkit, \textit{Resemblyzer}\footnote{https://github.com/resemble-ai/Resemblyzer},  to extract the speaker embedding from converted utterances of the same speaker and calculate the cosine similarity between them.  The greater the value of cosine similarity, the higher the homogeneity between different utterances. Based on the descriptions above, we employed two methods to match the utterances and calculate the cosine similarity between them. (1) We employed a randomized approach to match utterances converted by different face images of the same target speaker. To avoid chance, we shuffled the utterances 500 times and calculated the average cosine similarity between all pairs, which we referred to as the speaker homogeneity score by random matching (\textbf{SHR}). (2) In addition to the random matching, we also conducted one-to-one matching of all utterances converted from the same utterance to the same target speaker by different face images. We then averaged cosine similarity between these pairs to obtain the speaker homogeneity score by one-to-one matching (\textbf{SHO}).

Apart from homogeneity,  it is crucial for the voice characteristics converted from different target speakers to be diverse rather than uniform and indistinguishable. Similar to homogeneity, we obtained the speaker embedding of converted utterances from the same source speaker using different target speakers' face images by \textit{Resemblyzer} and calculated the cosine similarity between them.  We hypothesize that a lower similarity indicates higher diversity in the voice characteristics between target speakers. To measure speaker diversity, we also matched the utterance in two ways. (1) We randomly matched the converted utterances from the same source speaker to different target speakers and averaged the cosine similarity between them in 100 shuffles, which we referred to as the speaker diversity score by random matching (\textbf{SDR}). (2) We conducted a one-to-one matching of utterances converted from the same utterance to different target speakers and averaged cosine similarity between them to obtain the speaker diversity score by one-to-one matching (\textbf{SDO}).

When assessing the consistency between the voice characteristics and the corresponding face images, the gender attribute is the primary factor to consider.
Hence, we used the open source speech segments toolkit, \textit{inaSpeechSegmenter}\footnote{https://github.com/ina-foss/inaSpeechSegmenter}, to calculate gender accuracy (\textbf{GA}) for each converted utterance.  Specifically, a speech segmenter \cite{Doukhan2018open} was firstly used to discard segments that did not contain any speech. Next, the remaining speech segments were classify into either male or female using the convolutional neural networks. The gender of a converted utterance was consistent with the target speaker if all the speech segments in this utterance were classified as the gender of the target speaker.

For subjective evaluation, we adopted two mean opinion scores in terms of face-voice consistent degree (\textbf{MOS-FVC}) and speech naturalness (\textbf{MOS-SN}).  MOS-FVC was used to evaluate whether the face image and the voice characteristics were consistent with each other, e.g., a middle-aged man's face with the little girl's voice would be considered inconsistent. MOS-SN was used to quantitatively measure the naturalness of the converted voice. The listeners were asked to score each converted utterance on a scale from 1 (completely unnatural or completely inconsistent) to 5 (completely natural or completely consistent) for two metrics. 

\begin{figure*}[h] 
  \includegraphics[width=\textwidth]{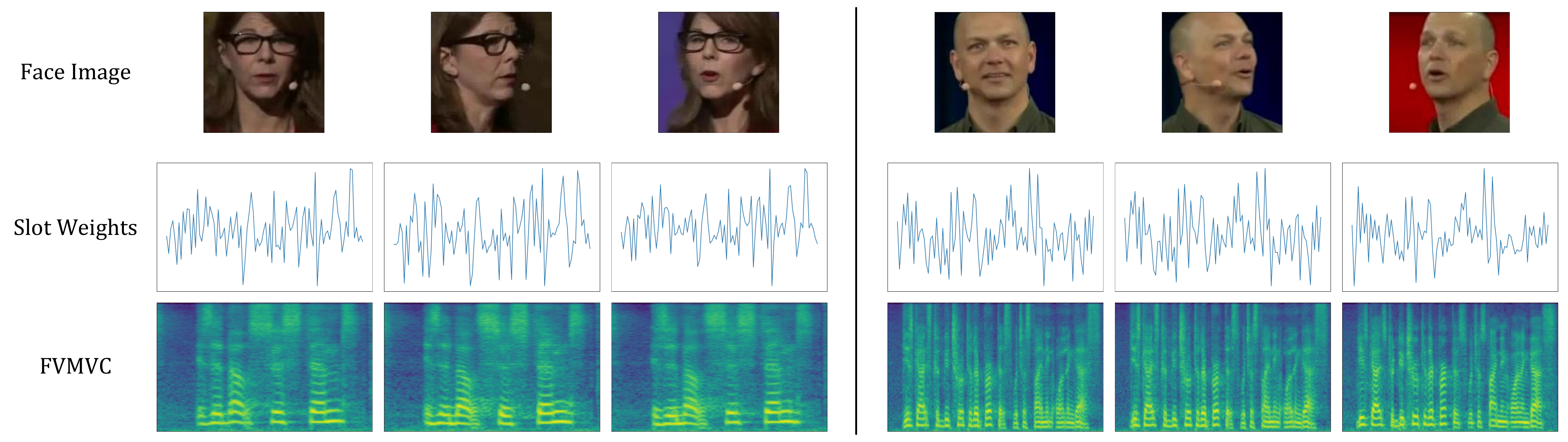}
  \caption{Target face images and their corresponding slots weights calculated by the MFVA module at the inference stage. The third row represents the mel-spectrograms of the converted utterance by FVMVC. }
  \label{fig3}
  \end{figure*}

\begin{figure}[h]  
  \includegraphics[width=0.45\textwidth]{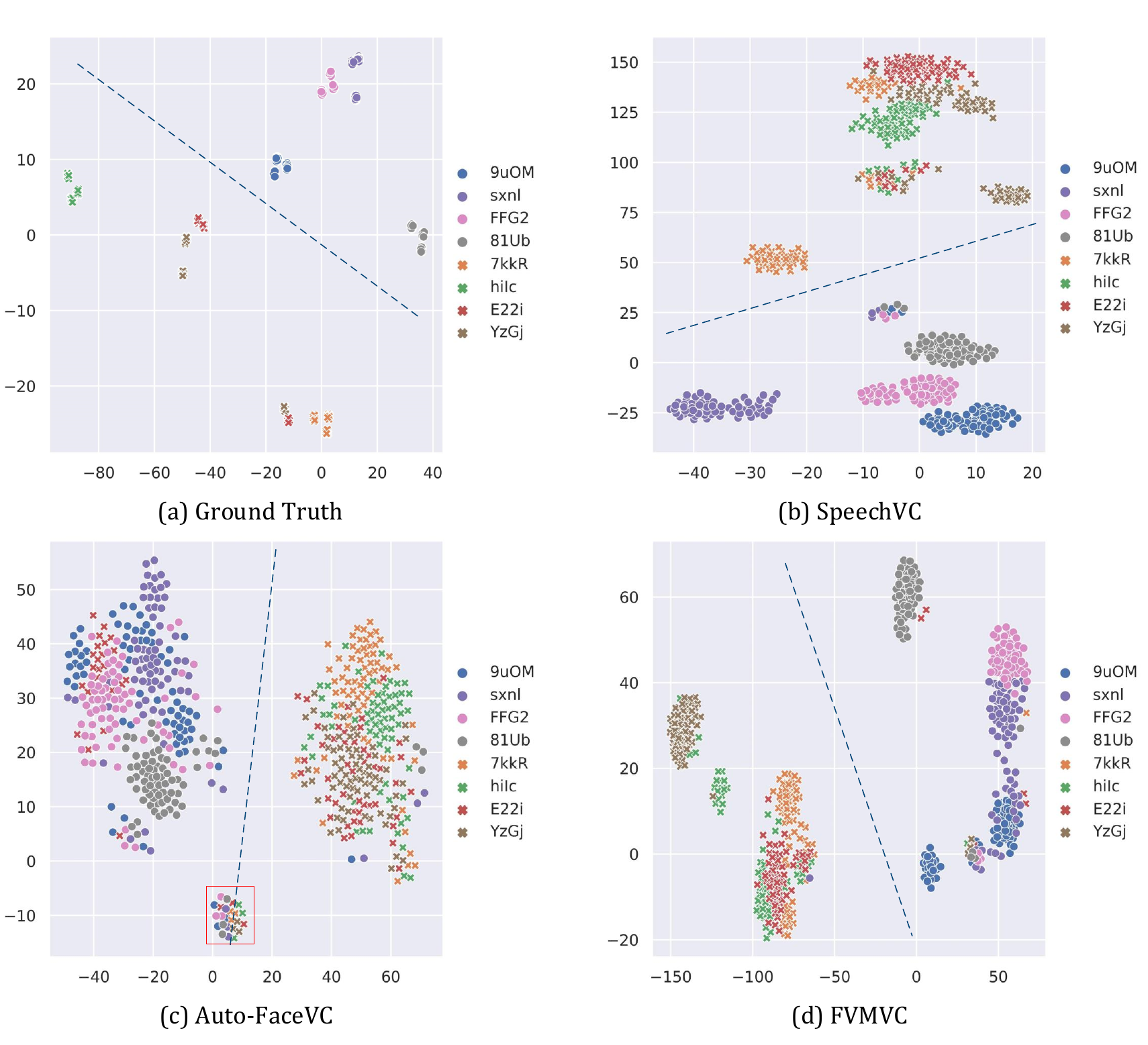}
  \caption{The t-SNE visualization of the speaker embeddings extracted from 576 utterances converted by different systems. Each point corresponds to a single utterance, with the colour of each point indicating the identity of the target speaker, ${\bullet}$ represents the male target speakers and ${\bm{\times}}$  represents female target speakers. }
  \label{fig4}
  \end{figure}

\subsection{Evaluation Results}
We chose 6 utterances from each of the 4 source speakers and randomly selected one face frame in 3 videos from each of the 8 target speakers for inference. Then we matched them pairwise and converted a total of 576 utterances for objective evaluation.  Two subjective metrics were evaluated on the Amazon Mechanical Turk platform\footnote{https://www.mturk.com/}. 20 converted utterances were randomly selected from each system and a total of 20 listeners participated in the test. All objective and subjective evaluation results are reported in Table~\ref{tab:compare}.

We can observe that the proposed \textbf{FVMVC} outperformed the \textbf{Auto-FaceVC} and \textbf{attentionCVAE} systems on all objective metrics significantly (${p<0.05}$ in paired t-tests). Compared with \textbf{Auto-FaceVC}, the slots in MFVA quantify the voice characteristics space, which makes the voice control via face images more homogeneous.  Additionally, our proposed \textbf{FVMVC} incorporates the MFVA module to alleviate the problem of over-smoothing, resulting in a more diverse range of voice characteristics.   In \textbf{attentionCVAE}, facial attributes can only provide limited information such as gender, age, and ethnicity.  As a result, while this method ensures relatively consistent voice characteristics and accurate gender, it also tends to generate very similar voices for different target speakers, resulting in a loss of diversity.  In addition, following the method described in \textbf{attentionCVAE} \cite{yang2022speaker}, we found that the facial attributes may vary across different face images of the same speaker, which can lead to heterogeneity among the voice characteristics of the same target speakers. With regard to GA, our proposed \textbf{FVMVC} has demonstrated a significant improvement, increasing from 90.85\% and 91.66\% in the two aforementioned methods to 97.91\%. The MOS-FVC has a strong correlation with the GA. In cases where the voice and the face image display a clear gender mismatch, the consistency score between them tends to be significantly low.  Since the three methods employed the same backbone, their performance in terms of MOS-SN is quite comparable.

We can observe that our proposed \textbf{FVMVC} performs better than \textbf{SpeechVC} with respect to SHR and SHO. This could be attributed to the presence of stable identity information in face images and several speaker-independent factors included in natural reference speech, such as prosody and emotions. These factors can influence the homogeneity between the converted utterances of the same target speakers.  In terms of SDR, SDO and MOS-FVC, our proposed method is less effective than \textbf{SpeechVC}, which is caused by the limited amount of voice characteristics information contained in the face image compared to the one contained in the natural reference speech. Additionally, for GA and MOS-SN, the performance of our proposed \textbf{FVMVC} and \textbf{SpeechVC} is essentially similar.

\begin{figure*}[h]
  \includegraphics[width=\textwidth]{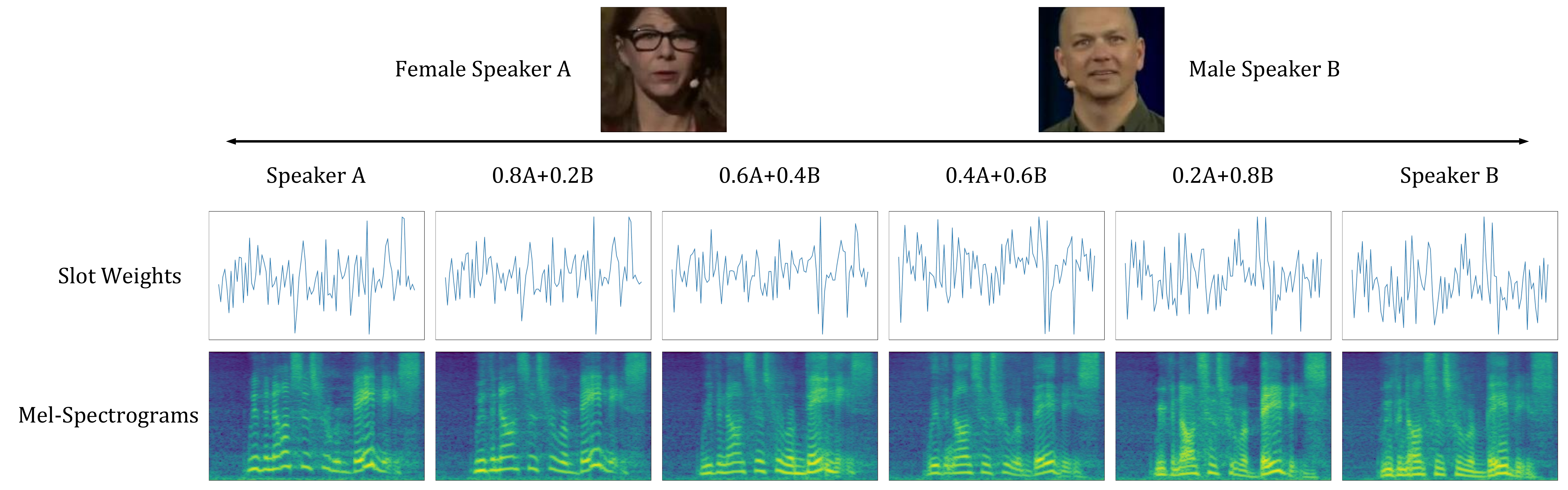}
     \caption{Voice characteristics interpolation by mixing the slot weights of a female speaker $A$ and a male speaker $B$.  From left to right, the slot weights of the male speaker increases sequentially. Specifically, ${0.6A+0.4B}$  means that we combine the slot weights of speaker $A$ and speaker $B$ with a weight of 6:4 to obtain a new recalled face embedding. The third row is the mel-spectrograms of the utterance converted by the corresponding slot weights.}
 \label{fig5}
  \end{figure*}
  
\subsection{Visual Analysis}
We utilized \textit{Resemblyzer} to extract speaker embeddings from the utterances generated by the \textbf{Ground Truth} and those converted by three other systems, i.e.,  \textbf{SpeechVC}, \textbf{Auto-FaceVC}, and \textbf{FVMVC}. We present their t-SNE \cite{chan2019gpu}  visualization in Figure \ref{fig4}. For the \textbf{Auto-FaceVC} system, some embedding clusters contained both male and female target speakers, as shown in the red box of Figure \ref{fig4}(c). This led to that a single converted utterance may contain both male and female voices in different segments. On the other hand, our proposed \textbf{FVMVC} model produced embeddings with a clear boundary between two genders, which further demonstrates the effectiveness of our method on GA and MOS-FVC metrics. In addition,  the embeddings of different target speakers overlapped a lot  for the \textbf{Auto-FaceVC} system. While similar to \textbf{SpeechVC},  our proposed \textbf{FVMVC} model had much less embedding overlap across target speakers, which also further confirms the better speaker diversity achieved by our method.

\subsection{Case Study}
We selected 2 target speakers' 6 face images taken from different perspectives for voice conversion, as shown in Figure \ref{fig3}. The first three columns belong to the first target speaker, and the last three columns belong to the second target speaker. The slot weights and corresponding mel-spectrograms of converted utterances based on the face images are visualized. 
We discover that the distributions of slot weights remain consistent across different face images of the same speaker, regardless of the angle or expression displayed in the images. This finding suggests that the speaker's facial features are of decisive importance in the process of aligning face and voice, and are minimally affected by external factors such as camera position, background, and other sources of noise.  As we can see from the third row in Figure \ref{fig3}, with the aid of stable recalled face embeddings, the mel-spectrograms converted by different face images exhibit a high level of uniformity.

Additionally, we attempted to achieve voice characteristics interpolation by manipulating the slot weights in the MVFA module. We chose a male and a female target speakers for creating new voices by interpolation, as depicted in Figure \ref{fig5}.
Specifically, we blended the slot weight of two face images with distinct weights to obtain the new recalled face embedding.  From left to right, as the slot weights of the male speaker $B$ increases, the voice characteristics gradually shift from female to male, and the fundamental frequencies gradually decreases.  This further validates the effectiveness of the MFVA module for face-based voice control.

  \subsection{Ablation Study}
\begin{table}[] 
\caption{Objective evaluation results of Ablation Studies. The definitions of all metrics can be found in Section \ref{metrics}.}
\label{tab:Abl1}
\begin{tabular}{lccccc}
\toprule
\multicolumn{1}{c}{Method} & SHR$\uparrow$    & SHO$\uparrow$    & SDR$\downarrow$    & SDO$\downarrow$    & GA$\uparrow$     \\ \midrule
\textbf{FVMVC}                              & \textbf{0.7313}& \textbf{0.8692} & \textbf{0.6188} & \textbf{0.6781} &\textbf{0.9791}   \\ \midrule
w/o Inter-speaker                    & 0.7301 & 0.8629 & 0.6262 & 0.6908 & 0.9444 \\
w/o MFVA                           & 0.7124 & 0.8257 & 0.6321 & 0.7111 & 0.9167 \\
w/o Pretraining                       & 0.7013 & 0.8113 & 0.6436 & 0.7140 & 0.8925 \\ \bottomrule
\end{tabular}
\end{table}

In this section, we conducted ablation experiments on our proposed \textbf{FVMVC} to explore the effectiveness of each module. As shown in Table~\ref{tab:Abl1}, we conducted experiments by removing the inter-speaker supervision, the MFVA module and the pretraining strategy from the proposed \textbf{FVMVC}, respectively.  The results show that when inter-speaker supervision was removed, the model's performance on SDR, SDO and GA decreased. This suggests that alleviating the inconsistency between training and inference phases can help to fit the recalled face embedding to the decoder, resulting in more diverse voice generation. 
After removing the MFVA module, the output of the face encoder was directly fed into the decoder without any constraints imposed by speaker embeddings. We find that the model's performance decreased in all aspects, highlighting the crucial role of alignment between face and voice in the model's performance. We also attempted to train the model from scratch without pretrained strategy. Our results show that the pretraining on the zero-shot VC task has a significant positive impact on the proposed \textbf{FVMVC} model. 

\section{Conclusion}
In this paper, we propose the FVMVC model to tackle a novel task of zero-shot FaceVC. The slots in the MFVA module act as a link between face and voice, promoting the performance of voice control based on face images of unseen speakers.  In addition, we have implemented a mixed supervision strategy to alleviate the long-standing issue of inconsistency between training and inference in VC tasks. As a result, based on the face images of newly coming speakers, the proposed FVMVC is able to generate a more consistent and diverse voice. 

As a future direction, we aim to explore the unified pre-training face-voice alignment,with a specific emphasis on voice control within text-to-speech, voice conversion, and singing synthesis tasks. Additionally, we plan to compile a comprehensive, large-scale video dataset featuring multiple speakers, ensuring its cleanliness and incorporating speaker details like age, gender, race, and physical appearance descriptions.

\clearpage
\bibliographystyle{ACM-Reference-Format}
\bibliography{facevc}


\end{document}